\documentstyle[aps,prl,multicol,psfig,floats]{revtex}

\begin{document}

\preprint{draft \today}

\twocolumn[
\hsize\textwidth\columnwidth\hsize\csname@twocolumnfalse\endcsname

\draft

\title{Mean-Field Analysis of a Dynamical Phase Transition
in a Cellular Automaton Model for Collective Motion}

\author{Harmen J. Bussemaker$^1$, Andreas Deutsch$^2$,
	Edith Geigant$^2$}

\address{${}^1$Institute for Physical Science and Technology,
	University of Maryland, College Park, Maryland 20742\\
	${}^2$Theoretical Biology, University of Bonn,
 	D-53115 Bonn, Germany	}

\date{Phys.\ Rev.\ Lett.\ 78, 5018--5021 (1997)}

\maketitle

\begin{abstract}
A cellular automaton model is presented for random walkers with
biologically motivated 
interactions favoring local alignment and leading to collective motion
or swarming behavior.
The degree of alignment is controlled by a sensitivity parameter,
and a dynamical phase transition exhibiting spontaneous breaking of
rotational symmetry occurs at a critical parameter value.
The model is analyzed using nonequilibrium mean field theory:
Dispersion relations for the critical modes are derived, and
a phase diagram is constructed.
Mean field predictions for the two critical exponents describing the 
phase transition as a function of sensitivity and density are obtained 
analytically.

\end{abstract}
]

\newpage

When in the course of evolutionary events it became possible for cells
to actively crawl and move towards more favorable habitats,
this led to an acceleration of evolutionary change.
Another important step was the development of social behavior,
manifested in cooperative motion of individual cells or organisms.
In particular, a change from independent crawling to cooperative motion is  
typical of life cycles in many microorganisms.

Moving cells can orient themselves by means of {\em indirect} physico-chemical 
signals like electrical fluxes or molecular concentration gradients;
the response of individual cells to such environmental information may
result in collective streaming behavior and swarm patterns.
Many models have been formulated along these lines of argument, 
all focusing on similar aspects of physico-chemical communication 
(see examples in Ref.~\cite{Alt97}).

Here we are interested in the implications of {\em direct}
communication between biological units (e.g. cells or birds).
Based on the assumption that the units have an inherent direction of motion,
and try to locally align with other units,
several microscopic models for swarming behavior have recently been proposed
\cite{Csahok95,Vicsek95,Alb96}.
These models can be viewed as itinerant 
$XY$-models that can be analyzed using renormalization group methods,
starting from a postulated equation of motion~\cite{Toner95}.

In this Letter we take a different approach.
We define a cellular automaton model~\cite{Deutsch96} that has the necessary
features to produce swarming behavior, while the discreteness
in time and space allows for relatively easy analysis.
We analyze our model {\em directly}, using an approximate mean-field kinetic
equation, and identify and derive dispersion relations for the various
collective modes.
An important question is how alignment is achieved, starting from a
random spatial distribution.
We show that swarm formation is associated with a continuous dynamical phase 
transition, occurring when a sensitivity parameter reaches a critical value.  
Spontaneous symmetry breaking leads to states with a global particle drift.  
The initial formation of patches is related to the fact that only at  
sufficiently  large
wavenumbers the density and longitudinal momentum modes merge to
form a pair of propagating sound modes.
We calculate the critical exponents governing the behavior of the average
drift velocity close to criticality.

The model we use is a lattice gas cellular automaton \cite{Doolen}
defined on a two-dimensional $L\times L$ square lattice with periodic 
boundary conditions.
Each node $\bf r$ can contain up to four particles
in different velocity channels corresponding to nearest neighbor vectors
${\bf c}_i=(\cos\phi_i,\sin\phi_i)$ with $\phi_i=\pi(i-1)/2$ and
$1 \leq i \leq 4$.
The state of the entire lattice at time $t$ is specified by the
occupation
numbers $s_i({\bf r},t)=0,1$ denoting the absence resp.\ presence of a
particle in the channel $({\bf r},{\bf c}_i)$.
The state of node $\bf r$ is denoted by
$s({\bf r},t)=\{s_i({\bf r},t)\}_{1\leq i\leq 4}$.

The evolution from time $t$ to time $t+1$ proceeds in two stages:
first an interaction step is performed during which the
preinteraction state $\{s_i({\bf r},t)\}$ is replaced by a
postinteraction state
$\{\sigma_i({\bf r},t)\}$ according to stochastic rules that are
applied to each node $\bf r$ independently;
the interaction step is followed by a propagation step during which
particles
move to nearest neighbor sites in the direction of their velocity, i.e.,
$s_i({\bf r}+{\bf c}_i,t+1)=\sigma_i({\bf r},t)$.

\begin{figure}[tb]
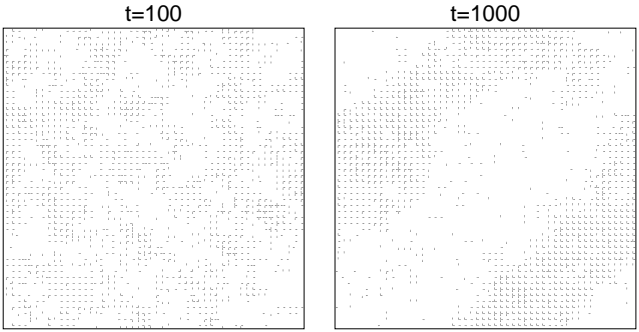

\centerline{
\psfig{file=fig1a.eps,width=40mm}
\hfill
\psfig{file=fig1b.eps,width=40mm}
}
\vspace{3mm}
\caption{
Swarming behavior in a cellular automaton model.
Shown are snapshots of the systems after 100 and 1000 time steps.
Parameters are: sensitivity $\beta=1.5$, system size $L=50$,
and average density $\bar\rho=0.8$.}
\label{fig:snapshots}
\end{figure}

To implement the local alignment interaction we define
\begin{equation}\label{D}
	{\bf D}({\bf r},t) = \sum_{p=1}^4 \sum_{i=1}^4
	{\bf c}_i s_i({\bf r}+{\bf c}_p,t),
\end{equation}
specifying the average flux of particles at the nearest neighbors
of node $\bf r$.
We require that the number of particles at each node,
$
	\rho({\bf r},t) = \rho[s({\bf r},t)]
	\equiv \sum_{i=1}^4 s_i({\bf r},t),
$
is conserved during interaction;
this implies that the spatially averaged density of particles per node
$\bar\rho$ is constant in time.
Let ${\bf J}(\sigma)=\sum_{i=1}^4 {\bf c}_i \sigma_i$ be the
particle flux immediately after interaction.
The transition probability from $s({\bf r},t)$ to
$\sigma({\bf r},t)$
in the presence of ${\bf D}({\bf r},t)$ is given by
\begin{equation}\label{A}
	A[s\to\sigma | {\bf D}] =
	\frac{1}{Z}\delta[\rho(\sigma),\rho(s)]
	\exp\left[ \beta {\bf D}\cdot{\bf J}(\sigma)\right],
\end{equation}
where the normalization factor $Z(\rho(s),{\bf D})$ is chosen
such that $\sum_\sigma A[s\to\sigma|{\bf D}]=1$ for all $s$.
The interaction rules are designed to minimize the angle between the
director field $\bf D$
and the postinteraction flux ${\bf J}(\sigma)$.
The sensitivity parameter $\beta$, playing the role of an inverse
temperature, controls the degree of local alignment:
for $\beta=0$ there is no alignment at all;
for $\beta\to\infty$ the two-dimensional inner product
${\bf D}\cdot{\bf J}(\sigma)$
--- and therefore the local alignment --- is maximized.
It will turn out that a dynamical phase transition occurs at a critical
value $\beta_c$ of the sensitivity.
Figure~\ref{fig:snapshots} shows the time evolution of an initially random
distribution for $\beta>\beta_c$.
The formation of locally aligned patches can clearly be observed.
There is some anisotropy due to the square lattice; it is however
straightforward to extend the model to the triangular lattice.
It is an interesting question whether the phase ordering kinetics shown in
Fig.~\protect\ref{fig:snapshots} can be described in terms of dynamical
scaling theory \cite{Langer}.

To analyze the behavior of the model we consider the time evolution of
a statistical ensemble of systems.
For technical details we refer to Ref.~\cite{Bussemaker96}, where
a model with only slightly different interaction rules
\cite{Alexander} yet entirely different behavior was analyzed.
In a mean-field description a central role is played by
the average occupation numbers
$f_i({\bf r},t) \equiv \langle s_i({\bf r},t)\rangle$.
It is assumed that at each time step just before interaction the
probability distribution is completely factorized over channels
$({\bf r}, {\bf  c}_i)$, so that the probability to find a microstate
$\{s_i({\bf r})\}$ at time $t$ is given by
$\prod_{\bf r} \prod_{i=1}^4 [f_i({\bf r},t)]^{s_i({\bf r})}
[1-f_i({\bf r},t)]^{1-s_i({\bf r})}$.
We denote the factorized average by $\langle\cdots\rangle_{\rm MF}$.
Replacing $\langle\cdots\rangle$ by $\langle\cdots\rangle_{\rm MF}$,
i.e., neglecting all correlations between occupation numbers,
we obtain a closed evolution equation for $f_i({\bf r},t)$:
the nonlinear Boltzmann equation,
\begin{equation}\label{nlbe}
	f_i({\bf r}+{\bf c}_i,t+1) = f_i({\bf r},t) + I_i({\bf r},t).
\end{equation}
Here the term
$I_i({\bf r},t)\equiv\langle\sigma_i({\bf r},t)-s_i({\bf r},t)\
\rangle_{\rm MF}$, taking values between $-1$ and $1$,
equals the average change in the occupation number of channel
$({\bf r},{\bf c}_i)$ during interaction.

\begin{figure}[b]
\psfig{file=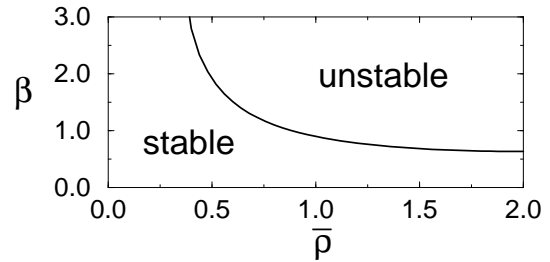,width=\columnwidth}
\caption{
Phase diagram for swarming model.
Shown are the regions of stable and unstable behavior,
as a function of sensitivity $\beta$ and average density $\bar\rho$.}
\label{fig:phase_diagram}
\end{figure}

It follows from the conservation of particle number, $\sum_i I_i=0$,
combined with the invariance of the interaction rules under discrete
rotations and translations that a possible solution to Eq.~(\ref{nlbe})
is $f_i({\bf r},t)=\bar f=\bar\rho/4$.
To assess the stability of this spatially homogeneous and stationary
solution with respect
to fluctuations $\delta\!f_i({\bf r},t) = f_i({\bf r},t) - \bar f$ we
linearize Eq.~(\ref{nlbe}), perform a Fourier transformation,
$\delta\!f_i({\bf k},t) = \sum_{\bf r} e^{-i{\bf k}\cdot{\bf r}}\
\delta\!f_i({\bf r},t)$ \cite{i-footnote},
and obtain
\[
	\delta\!f_i({\bf k},t+1) \simeq
	\sum_{j=1}^4 \Gamma_{ij}({\bf k}) \delta\!f_j({\bf k},t).
\]
The mean-field or Boltzmann propagator $\Gamma({\bf k})$ describes how a
small perturbation around a spatially uniform state evolves in time.
It is given by
\[
	\Gamma_{ij}({\bf k}) = e^{-i{\bf k}\cdot{\bf c}_i}
	\left[ \delta_{ij} + \sum_{p=0}^4 e^{i{\bf k}\cdot{\bf c}_p}
	\ \Omega^p_{ij} \right],
\]
with ${\bf c}_0\equiv0$ and
$\Omega^p_{ij}=\partial I_i({\bf r},t)/\partial f_j({\bf r}
+{\bf c}_p,t)|_{\bar f}$.
It can be shown that $\delta_{ij}+\Omega^0_{ij}=1/4$ for all $i,j$;
this is a consequence of the fact that the outcome $\sigma({\bf r})$
of an interaction step only depends on $s({\bf r})$ through
$\rho({\bf r})$ (see Eq.~(\ref{A}) and Ref.~\cite{Bussemaker96}).
For $1\leq p\leq 4$ the elements $\Omega^p_{ij}\equiv\omega_{ij}$
do not depend on $p$, as can be seen from the definition of ${\bf D}$
in  Eq.~(\ref{D}).
We note that $(\omega)_{ij}$ is a cyclic matrix whose first row has the
structure $(\alpha+\gamma, -\gamma, -\alpha+\gamma, -\gamma).$
To determine $\alpha(\beta,\bar\rho)$ and $\gamma(\beta,\bar\rho)$
for given values of the sensitivity $\beta$ and the average density
$\bar\rho$ we evaluate the expression
(this is done numerically because of the highly nonlinear
dependence on $f_i$
and $\beta{\bf D}$, combined with the large number of terms)
\begin{eqnarray*}
	\omega_{ij} &=&
	\sum_{ \{s({\bf r}+{\bf c}_p)\}} \sum_{\sigma({\bf r})}\
	(\sigma_i({\bf r})-s_i({\bf r}))
	\frac{s_j({\bf r}+{\bf c}_1) - \bar f}{\bar f(1-\bar f)} \\
	&& \times \  A[s\to\sigma|{\bf D}(\{s({\bf r}+{\bf c}_p)\})]
	\prod_{p'=0}^4 F(s({\bf r}+{\bf c}_{p'})),
\end{eqnarray*}
where $F(s)=\prod_{i=1}^4 \bar f^{s_i} (1-\bar f)^{1-s_i}$
is the factorized distribution.
Note that the expression for $\omega_{ij}$
does not depend on $\bf r$ since it
represents a derivative evaluated in a spatially uniform state.

We first investigate the stability of the spatially uniform state,
i.e.\ ${\bf k}=0$.
It can be seen that the propagator $\Gamma_{ij}({\bf k}=0)$
has an eigenvalue $\lambda_1=1$ with corresponding eigenvector
$e_1=(1,1,1,1)$,
reflecting the fact that the total density is conserved.
Furthermore there is a twofold degenerate eigenvalue
$\lambda_{x,y}=8\alpha$
with an eigenspace spanned by $e_x=(1,0,-1,0)$ and $e_y=(0,1,0,-1)$,
corresponding to the $x$- and $y$ components of the total particle flux.
The remaining eigenvector $e_{x^2-y^2}=(1,-1,1,-1)$ has eigenvalue
$\lambda_{x^2-y^2}=16\gamma$, corresponding to the difference
between the
number of horizontally and vertically moving particles.
Numerically $\gamma$ is found to be about two orders of magnitude
smaller
than $\alpha$, so that the onset of instability of the
homogeneous state is
determined by the condition $\lambda_{x,y}=1$.
The location of the critical line in the $(\beta,\bar\rho)$
parameter plane is shown in Fig.~\ref{fig:phase_diagram},
which was obtained by numerically solving the equation
$\alpha(\beta,\bar\rho)=1/8$.

\begin{figure}[t]
\psfig{file=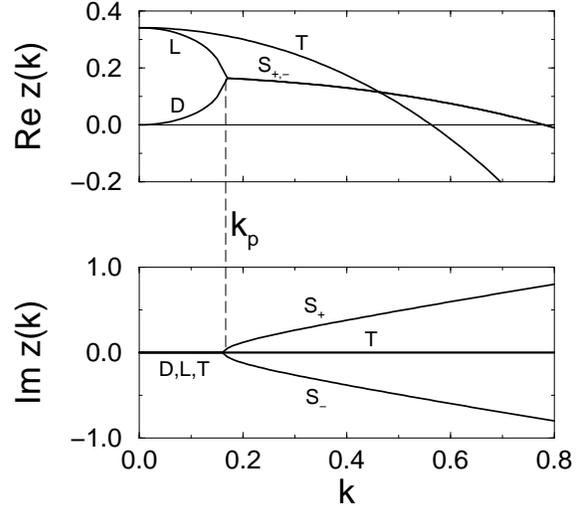,width=\columnwidth}
\caption{
Eigenvalue spectrum for $\bar\rho=1.6$, $\beta=1.5$,
and ${\bf k}/\!\!/\hat{\bf x}$.
Density (D), longitudinal (L) and transverse (T) momentum, and
sound ($S_\pm$) modes are indicated.
The stable mode that has eigenvector $e_{x^2-y^2}$ at ${\bf k}=0$ is
not shown.
}
\label{fig:spect}
\end{figure}

To see if in addition to the emergence of a global drift we can explain
the formation of spatial structure in terms of the eigenvalue spectrum,
we study the case ${\bf k}\neq 0$.
It is convenient to work with $z({\bf k})=\ln\lambda({\bf k})$
so that excitations behave as
$\delta\!f({\bf r},t)\sim\exp[z({\bf k})t+i{\bf k}\cdot{\bf r}]$.
Unstable modes have ${\rm Re}\,z({\bf k})>0$ while stable modes have
${\rm Re}\,z({\bf k})<0$.
An imaginary part of $z({\bf k})$ indicates that the mode has a nonzero
propagation velocity $v({\bf k})={\rm Im}\;z({\bf k})/|{\bf k}|$.
Figure~\ref{fig:spect} shows that the fastest growth occurs at
${\bf k}=0$.
For ${\bf k}\neq0$ the degeneracy of $\lambda_{x,y}$ is
lifted, and it is then the transverse velocity
(i.e., perpendicular to $\bf k$) that grows fastest.
At $|{\bf k}|=k_p$, with $k_p=k_p(\hat{\bf k},\bar\rho,\beta)$,
where $\hat{\bf k}$ is the unit vector in the direction of $\bf k$,
the density and longitudinal velocity modes merge
to form a pair of propagating sound-like modes, with
${\rm Im}\;z({\bf k})\neq 0$, and traveling in the directions
$\pm\hat{\bf k}$.
Thus, traveling waves cannot occur on spatial scales larger than
$2\pi/k_p$, which may explain the length scale for short times
of the spatial structure shown in Fig.~\ref{fig:snapshots}.

Our mean-field stability analysis illuminates the nature of the
observed phase transition.
An appropriate order parameter is the spatially averaged velocity,
\[
	\bar \mu(t) = \frac{1}{L^2}\left|\sum_{\bf r}
	\sum_{i=1}^{4}{\bf c}_{i}s_{i}({\bf r},t)\right|,
\]
which takes values between 0 and 1.
For $\beta<\beta_c$ we have $\bar\mu=0$.
When the sensitivity parameter $\beta$ reaches its critical value,
this ``rest'' state becomes unstable, leading to a breaking of rotational
symmetry, and a stationary state where $\bar\mu\neq0$.

\begin{figure}[t]
\psfig{file=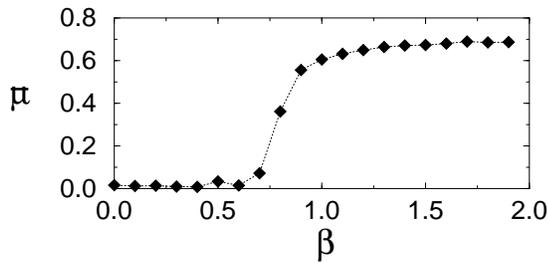,width=\columnwidth}
\caption{
Mean velocity $\bar\mu$ versus sensitivity $\beta$.
Obtained from simulation of $L=50$ system at averaged density
$\bar\rho=1.6$, after $t=1000$ time steps.}
\label{fig:mu}
\end{figure}

We have compared the results of our stability analysis with 
computer simulations.
Fig.~\ref{fig:mu} shows $\bar\mu$ versus $\beta$ for averaged density
$\bar\rho=0.4$.
There is an abrupt change in $\bar\mu$ at $\beta\simeq0.7$, which agrees
well with the prediction $\beta_c=0.67$ obtained from our stability
analysis.

A discussion of the question whether the transition is first order or
continuous is only meaningful if we consider the limit $t\to\infty$,
the analogue of the thermodynamic limit $L\to\infty$.
For $\beta<\beta_c$ all modes are stable and we have $\bar\mu(t\to\infty)=0$.
To determine the behavior of $\bar\mu$ for $\beta>\beta_c$ we consider
spatially homogeneous and stationary solutions to the nonlinear Boltzmann
equation (\ref{nlbe}),
i.e.\ $f_i({\bf r})=f_i$ and $I_i=0$.
Knowing that the ``rest'' solution, $f_i=\bar f=\bar\rho/4$, is stable for
$\beta<\beta_c(\bar\rho)$, we expand around the critical point
$(\bar\rho,\beta_c)$:
\begin{eqnarray*}
	I_i(\bar\rho+\Delta\rho,\beta_c+\Delta\beta) &=& \sum_k
	\left(\bar\Omega_{ik} + \frac{\partial\bar\Omega_{ik}}{\partial\beta}
	\Delta\beta \right) \delta\!f_k \\
	&& \mbox{} \hspace{-25mm}
	+ \frac{1}{2}\sum_{k_1 \leq k_2} \bar\Omega_{ik_1k_2}
	  \delta\!f_{k_1} \delta\!f_{k_2} + \ldots
\end{eqnarray*}
where $\bar\Omega_{ik_1\cdots k_n}=(\partial/\partial\!f_{k_1})
\cdots (\partial/\partial\!f_{k_n}) I_i$ \cite{Omega-footnote}.

We use a particular parametrization for ``drift'' solutions along the
x-axis:
$\delta\!f_1-\delta\!f_3=\bar\mu$, $\delta\!f_2=\delta\!f_4$, and
$\delta\!f_1+\delta\!f_2+\delta\!f_3+\delta\!f_4=\Delta\rho$.
Utilizing $I_1=I_2=0$, together with the symmetry properties of the
expansion
coefficients $\bar\Omega_{ik_1\cdots k_n}$ and the fact that at the critical
point all three vectors 1, $c_x$, and $c_y$ are zero eigenvectors of
$\bar\Omega_{ik}$, we can eliminate $\{\delta\!f_i\}$ and
for small $\Delta\rho$ and $\Delta\beta$ obtain the following
equation of state:
\begin{equation}\label{eos}
	(c_\beta \Delta\beta + c_\rho \Delta\rho) \bar\mu - \bar\mu^3 \simeq 0.
\end{equation}
Here $c_\beta$ and $c_\rho$ are positive constants that depend on
the expansion
coefficients $\bar\Omega_{ik_1\cdots k_n}$.

Consider now the case $\Delta\beta\neq0$, $\Delta\rho=0$.
The solution $\bar\mu=0$ is
stable only for $\beta<\beta_c$, or $\Delta\beta<0$.
From Eq.~(\ref{eos}) we see that for $\Delta\beta>0$ there is an
additional, stable solution $\bar\mu\sim\sqrt{\beta}$.
Thus for the critical exponent $\beta'$ \cite{exp-footnote}
defined by $\bar\mu\sim(\beta-\beta_c)^{\beta'}$ in Ref.~\cite{Vicsek95}
we find $\beta'=\case{1}{2}$.
A different exponent $\delta$, defined by
$\bar\mu\sim(\rho-\rho_c)^\delta$,
governs the behavior for $\Delta\rho\neq0$, $\Delta\beta=0$.
From Eq.~(\ref{eos}) we obtain $\delta=\case{1}{2}$.

The essential elements of our analysis
are the ``hydrodynamic'' variables density and velocity.
Therefore we expect that many of our predictions
--- including the value of the critical exponents ---
should also apply to the continuum swarming model of
Ref.~\cite{Vicsek95}.
From a coarse-grained point of view our model and that
of Ref.~\cite{Vicsek95} are equivalent.
In particular the noise parameter $\eta$ in Ref.~\cite{Vicsek95}
plays a role analogous to $1/\beta$ in our model.

Our analysis confirms the numerical finding of Ref.~\cite{Vicsek95}
that the phase transition is continuous, but is in conflict with the
results of Ref.~\cite{Csahok95}.
The exponents $\beta'$ and $\delta$ have been measured in computer
simulations \cite{Vicsek95}.
The measured value $\beta'=0.45\pm0.07$ is in agreement with our
mean-field prediction $\beta'=\case{1}{2}$.
In the case of $\delta$ however there is a significant deviation between
the measured value $\delta=0.35\pm0.06$ and the mean-field result
$\delta=\case{1}{2}$.
The fact that at a mean-field level $\beta'$ and $\delta$ are equal
supports a claim  made by the authors of Ref.~\cite{Vicsek95} that the
observed difference between the measured values of the two exponents may
be due to finite-size effects.

Our model has an interesting  biological interpretation since the dynamical  
phase transition suggests two possible scenarios for a change from  
non-cooperative to cooperative behavior. On one hand, genetically caused  
minor microscopic effects on receptor properties of interacting cells  
influencing their sensitivity can have severe macroscopic implications with  
respect to swarming if they occur close to criticality (cf.~Fig.~2). On the  
other hand, a transition from the stable into the unstable region can also  
be achieved by simply increasing cell density (cf.~Fig.~2). This result  
provides a possible clue to explain the behavioral change between 
non-cooperative and cooperative stages in individual life cycles of some 
bacteria and amoebae in which a reproductive feeding phase of individually 
moving cells is followed by social (coordinated) aggregation.
Other models for complex bacterial pattern formation, including vortex and 
colony organization, have also been proposed 
(see Refs.~\cite{Ben96,Ben97} and references therein).

Finally, we want to stress that the methods employed here can easily be 
adapted to gain theoretical insight in the behavior of a wide range of
biologically motivated cellular automaton models, that so far have mainly
been analyzed by observation of simulation outcomes~\cite{other-models}.

The present research in part was supported by
Sonderforschungsbereich 256
(``Nonlinear partial differential equations'') and a NATO grant
(``Modelling the actin dynamics in cellular systems'').
Valuable interactions and discussions with Matthieu Ernst (Utrecht)
and Wolfgang Alt and Michael Stoll (Bonn) are highly appreciated.


\begin{references}
\bibitem{Omega-footnote}
The symbol $\bar\Omega_{ik}$ is related to $\Omega^p_{ik}$ by
$\bar\Omega_{ik}=\sum_{p=0}^4 \Omega^p_{ik}$.
\bibitem{exp-footnote}
We use the symbol $\beta'$ instead of $\beta$ for the exponent to
avoid confusion with the sensitivity parameter $\beta$.
\end{references}
\end{document}